\documentstyle[preprint,aps]{revtex}

\tightenlines
\begin{document}

\draft
\preprint{}
\title{Consistent Prediction for Direct CP Violation and
$\Delta I = 1/2$ Rule  }
\author{ Yue-Liang  Wu  }
\address{Institute of Theoretical Physics, Academia Sinica, \\
 P.O. Box 2735, Beijing 100080, P.R. China }
 \date{Plenary Talk at ICFP2001, Zhang-Jia-Jie}
\maketitle

\begin{abstract}
The theoretical status of direct CP violation
$\varepsilon'/\varepsilon$ is briefly reviewed. Special attention
is paid to the recent new consistent
 predictions for both the ratio $\varepsilon'/\varepsilon$
and the $\Delta I = 1/2$ rule within the standard model. In
 particular, two matching conditions resulting from the matching
 between the QCD and chiral perturbation theory(ChPT), and also some
algebraic relations of chiral operators are found to be very
useful. It is of interest that the new predictions are no longer
sensitive to the strange quark mass, and are also renormalization
scale and scheme independent in the leading QCD and chiral loop
approximation with large
 $N_c$ approach. The new prediction for the direct CP violation with the value
$\varepsilon'/\varepsilon=(20\pm 4)\times 10^{-4}
 [Im\lambda_t /1.2\times 10^{-4}]$ is consistent
with the most recent experimental results reported by the NA48 and KTeV groups.
\end{abstract}


\section{Introduction}

  One of puzzles in particle physics is the origin and mechanism of CP violation\cite{LW1}.
  In the standard model (SM), CP violation is described by the Kobayashi-Maskawa (KM)
  \cite{KM} phase in the Cabbibo-Kobayashi-Maskawa quark mixing matrix.
  Such a phase can arise from the explicit CP
  violation imposed in the Yukawa coupling constants of SM or originate
  from spontaneous CP violation\cite{TDL} in some extended models of SM.
  One of the simplest models is the extension of SM with two Higgs
  doublets (S2HDM) \cite{WW} motivated from spontaneous CP violation. In such
  an S2HDM, rich CP-violating sources can be induced from a single relative CP-violating phase
  of vacuum. These sources have been classified into four types\cite{WW}:
  (i) induced single KM phase $\delta_{KM}$ or Wolfenstein parameter $\eta$\cite{LW};
  (ii)  induced new type of flavor-dependent
  CP-violating phases $\delta_{f_{i}}$ via charged and neutral scalar interactions;
  (iii) induced super-weak type CP violation\cite{LW2} via flavor-changing scalar interactions;
  (iv) induced neutral scalar-pseudoscalar mixing CP violation. When the source of
  CP violation is dominated by the induced KM phase, the phenomena of CP violation
  in S2HDM  would be similar to the SM.

      To make a consistent prediction for the direct CP-violating parameter
  $\varepsilon'/\varepsilon$ caused by the KM CP-violating phase, it is necessary to
  understand simultaneously the longstanding puzzle of the $\Delta I = 1/2$ rule in
  the kaon decays as they involve the long-distance evolution of common hadronic matrix
  elements, where the low energy dynamics of QCD shall play a crucial role for a
  consistent analysis. During the past few years, both theoretical and experimental
  efforts on direct CP violation in the kaon decays have been made important progresses.
  On the experimental side, two improved new experiments\cite{NA48,KTEV} with higher
  precision have reported results which are consistent each other at the $1-\sigma$ level.
  On the theoretical side, several groups\cite{g1,g2,g3,g4,g5,g6,g7,g8,g9,g10,g11,g12,g13}
  have made detailed calculations for the ratio $\varepsilon'/\varepsilon$. Recently,
  there have been some interesting developments\cite{g1} which are
  mainly based on QCD of quarks and chiral perturbation theory (ChPT)\cite{CHPT0,CHPT}
  at low energies for mesons. Consequently, it has reached an agreement between
  the experimental results and the theoretical predictions.

   To reach a consistent theoretical prediction for the direct
  CP-violating parameter $\varepsilon'/\varepsilon$, it must satisfy, at least, the
  following simple criteria: (i) the prediction should be renormalization scale independent;
  (ii) the $\Delta I = 1/2$ rule can be reproduced.  To satisfy the first criteria (i),
  one should be able to solve the matching problem between perturbative QCD used to calculate
  the Wilson coefficient functions $c_i(\mu)$ and effective theories applied to evaluate
  the hadronic matrix elements of operators $<|Q_i|>(\mu)$. To reach the second one (ii),
  the effective theory should well describe the low energy dynamics of QCD in weak kaon decays.
  One of such attractive effective theories is the chiral perturbation
  theory (ChPT) inspired from $1/N_c$ expansion\cite{THOOFT,EW},
  which has actually provided a successful
  description on many processes\cite{CHPT}. In ref. \cite{g1},
  we have further shown that the ChPT with a functional cutoff
  momentum scheme can lead to a consistent prediction for the direct CP-violating
  parameter $\varepsilon'/\varepsilon$ and the $\Delta I =1/2$ rule.

\section{Brief Review on Theoretical Status}

  The present status of theory versus experiment is summarized as follows:
\\
  \begin{tabular}{|c|c|}
    \hline
    References & $\varepsilon'/\varepsilon$ $[10^{-4}]$ \\ \hline
    World Average & $17.2 \pm 1.8$ \\ \hline \hline
    NA48\cite{NA48} & $15.3 \pm 2.6$ \\ \hline
    KTeV\cite{KTEV} & $20.7\pm 2.8$ \\ \hline \hline
    Beijing\cite{g1} &  $20 \pm 4\pm 5 $\  \\ \hline
    Dortmund\cite{g2} & $ 6.8\rightarrow 63.9\ $ (S) \\ \hline
    Dubna\cite{g3} & $-3.2\rightarrow 3.3 $\ (S) \\ \hline
    Granada-Lund\cite{g4} & $ 34\pm 18 $\\ \hline
    Montpellier\cite{g5} & $ 4\pm 5 $\\
                & $\le 22\pm 9$ \\ \hline
    Munich\cite{g6} & $9.2^{+6.8}_{-4.0}$\ (G) \\
           &  $1.4\rightarrow 32.7$\ (S) \\ \hline
    Roma\cite{g7} & $8.1^{+10.3}_{-9.5} $\ (G) \\
         &  $-13.0\rightarrow 37.0$\ (S) \\ \hline
    Taipei\cite{g8} &  $ 7 \rightarrow 16 $ \\ \hline
    Trieste\cite{g9} & $22 \pm 8 $\ (G) \\
            &  $9\rightarrow 48$\ (S) \\ \hline
    Valencia\cite{g10} & $17\pm 9$ \\ \hline
  \end{tabular}
  \\
\\
  It is seen that great theoretical efforts
  have been made by many groups as a consistent prediction for
   $\varepsilon'/\varepsilon$ and $\Delta I = 1/2$
  rule  play an important role for understanding the origin and mechanism of CP violation
  as well as the low energy dynamics of QCD.
  Where different approaches have been adopted to obtain
 the above theoretical results:

  {\bf Beijing's\cite{g1}:}  The ChPT inspired from $1/N_c$ expansion
  has been adopted to calculate the chiral loop contributions. The claculating scheme
  is based on the one with cutoff momentum $M$ suggested by Bardeen, Buras and
  Gerard \cite{BBG}, but it is generalized by taking the cutoff momentum as the function
  of the renormalization scale $\mu$, i.e., $M = M(\mu)$, instead of naively identifying
  the cut-off momentum to the QCD running scale $\mu$. It is such a simple extention that
  enables us to make the new precitions for both the direct CP-violating
  parameter $\varepsilon'/\varepsilon$  and the $\Delta I =1/2$ rule
  be renormalization scale and scheme independent as well as be consistent with the experimental data
  The new predictions also nicely improved our previous ones presented in ref.\cite{YLW1}.

  {\bf  Dortmund's\cite{g2}:}   Chiral loops are calculated by separating the
   factorized and non-factorized diagrams\cite{DD}. The $\Delta I =1/2$ rule may be
   reproduced\cite{DDI}.
  While the matching procedure needs to be improved as the presence of the quadratic cutoff
   may induce a matching scale instability.

  {\bf  Dubna's\cite{g3}:} Chiral loops are regularized via the heat-kernel
   method in the ENJL framework up to $O(p^6)$. The
   renormalization scheme dependence needs to be improved. A
   phenomenological fit of the $\Delta I =1/2$ rule also results in a big
   uncertainty on the matrix element $\langle Q_6 \rangle$.

  {\bf  Granada-Lund's\cite{g4}:}  The X-boson method within the ENJL model is used
   to improve the matching
   between long- and short-distance components. The $\Delta I =1/2$ rule can be
   reproduced. Both renormalization scale and scheme dependences
   have been improved. While Non-FSI chiral corrections need to
   be further considered.

  {\bf  Montiepeller's\cite{g5}:} Effects to the matrix element
  $\langle Q_6 \rangle$ from the $\bar q q$ component of the scalar meson
  was considered by using  QCD sum rule.
  The question is how to explicitly separate the $\bar q q$ from gluon
   component of the scalar meson.

  {\bf  Munich's\cite{g6}:} A phenomenological $1/N_c$ approach is adopted, where some of
   the matrix elements were obtained by fitting the $\Delta I =1/2$ rule
   at $\mu = m_c = 1.3$ GeV. The matrix elements $\langle Q_6
   \rangle$ and $\langle Q_8 \rangle_2$ relevant to the direct
   CP-violating parameter $\varepsilon'/\varepsilon$ remain
   undetermined and are taken around their leading $1/N_c$ values.

  {\bf  Roma's\cite{g7}:}  Present lattice results remain unreliable as
  large renormalization uncertainties at the matching scale
   between the lattice and continuum results. The present lattice
   calculations can only use the lowest order chiral perturbation theory to
   evaluate the $K\rightarrow \pi\pi$ amplitude. The recent result
   for  $\varepsilon'/\varepsilon$ was estimated by taking $B_6$ to
   be the VSA result varied by a $100\%$ error.

  {\bf  Taipei's\cite{g8}:} Effective Hamiltonian approach was adopted in conjunction with
   generalized factorization for hadronic matrix elements. The
   non-factorizable effects were considered by introducing
   phenomenological parameters. Some assumptions are needed to fix
   the phenomenological parameters.

   {\bf Trieste's\cite{g9}:}  The chiral quark model has been used to evaluate the
   hadronic matrix elements which were assumed to be matched to the NLO
   short-distance Wilson coefficients at $\mu = 0.8$ GeV.
   The $\Delta I =1/2$ rule can be reproduced at that point. The problem of
   renormalization scale and scheme dependence needs to be improved.

   {\bf Valencia's\cite{g10}:}  A dispersive analysis \'{a} la Omn\'{e}s is used to
   obtain the FSI effects relative to the leading $1/N_c$
   amplitudes. The FSI effects alone cannot reproduce the $\Delta I =1/2$ rule.
   More comments on this approach may be found in ref.\cite{MR}.

 In most of the approaches, the matching procedure remains the main problem.
 This is because the energy scale $M$ of long-distance operator evolution from meson loops
 must in general be smaller than the chiral symmetry breaking scale
 $\Lambda_{f}$, i.e., $M < \Lambda_{f}\sim 1 $ GeV, while the energy scale
 $\mu$ of the short-distance operator evolution from perturbative QCD should
 be above the confining scale, i.e., $\mu >$ 1 GeV. Naively identifying the scale $M$ in ChPT
 to the scale $\mu$ in perturbative QCD may become inappropriate. In the following,
 we will focus our discussions on the recent new consistent predictions  for
 the direct CP-violating parameter $\varepsilon'/\varepsilon$ and
 the $\Delta I =1/2$ rule presented in ref.\cite{g1}.

\section{Low Energy Dynamics of QCD}

 To treat the low energy dynamics of QCD, our starting points are mainly based on
 the following basic considerations:

\begin{itemize}
\item In the large $N_c$ limit but with the combination $\alpha_s N_c \equiv \alpha_0$
       being held fixed. The QCD loop corrections which are proportional to
       $\alpha_s$ are then corresponding to a large $N_c$ expansion, $\alpha_s \sim 1/N_c$
       \cite{THOOFT}.
\item Chiral flavor symmetry $SU(3)_L\times SU(3)_R$ is supposed to be broken
       dynamically due to attractive gauge
       interactions, namely the chiral condensates $<\bar{q}q> $
        exist and lead to the Goldstone-like pseudoscalar mesons $\pi$, $K$,
       $\eta$. The chiral symmetry breaking scale $\Lambda_{f}$ is characterized
       by the condensate, $\Lambda_{f} \simeq 4\pi \sqrt{-2<\bar{q} q> /r } \sim 1$ GeV
        with $r = m_{\pi_0}^{2}/\hat{m}$ ($\hat{m} = (m_u + m_d)/2 $).
\item The low energy dynamics of QCD in large $N_c$ limit is considered to be described
       by the chiral Lagrangian. The ChPT is going to be treated as a cut-off effective field
       theory which may be regarded as a consistent theory in a more general sense\cite{GW} .
       The cut-off momentum $M$ is expected to be below the chiral symmetry breaking scale
       $\Lambda_{f}$.
\item The chiral meson loop contributions are characterized by the powers of
       $p^{2}/\Lambda_{f}^{2}$ with $\Lambda_{f}= 4\pi f$ .
       Here $f^{2} \simeq -2<\bar{q} q> /r \sim N_c$ is at the leading $N_c$ order and fixed by
       the $\pi$ decay coupling constant $f\sim F_{\pi}$.
       Thus the chiral meson loop contributions are also corresponding to a
       large $N_c$ expansion of QCD, $p^{2}/\Lambda_{f}^{2} \sim 1/N_c\sim \alpha_s$. Therefore
       both chiral loop and QCD loop contributions must be matched each other, at least in
       the sense of large $N_c$ limit. Thus the final physical results should be
       independent of the renormalization scale and the calculating schemes.
 \item The cut-off momentum $M$ of loop integrals is in general taken to be
       a function of $\mu$, i.e., $M \equiv M(\mu)$ \cite{g1},
       instead of naively identifying it to
       the renormalization scale $\mu$ appearing in the perturbative QCD in large $N_c$ limit.
       $M \equiv M(\mu)$ may be regarded as a functional cut-off momentum,
       its form is determined by the matching between the Wilson coefficients of QCD
       and hadronic matrix elements evaluated via ChPT.
  \end{itemize}

  In a word, we are going to treat the ChPT with functional cut-off momentum $M(\mu)$ scheme,
 as a low energy effective field theory of QCD in the large $N_c$ approach.

\section{Chiral Representations and Relations of Operators }

 The general form of the chiral Lagrangian is constructed based on the chiral flavor
 symmetry $SU(3)_L\times SU(3)_R$ and can be expressed
 in terms of the expansion of the momentum $p$ and the quark mass $m_q$
 to the energy scale $\Lambda_{\chi}= O(1)$ GeV at which nonperturbative effects
 start to play the rule. For our purpose, here we only use the chiral
 Lagrangian which is relevant to the $K\rightarrow \pi\pi$
 decays (for the most general one, see ref.\cite{CHPT})
  \begin{eqnarray}
{\cal L}_{eff}&=&\frac{f^2}{4}\{ \  tr(D_{\mu} U^{\dagger}
D^{\mu} U ) + \frac{m_{\alpha}^2}{4N_c} tr( \ln U^{\dagger} -\ln U
)^2 +r\ tr({\cal M} U^{\dagger}+U{\cal M}^{\dagger} ) \nonumber \\
& & + r \frac{\chi_5}{\Lambda_{\chi}^2} tr\left( D_{\mu}
U^{\dagger} D^{\mu}  U({\cal M}^{\dagger} U + U^{\dagger}{\cal M}
\right) \\
&& + r^2 \frac{\chi_8}{\Lambda_{\chi}^2} tr\left( {\cal
M}^{\dagger} U{\cal M}^{\dagger} U+{\cal M} U^{\dagger}{\cal M}
U^{\dagger} \right) + r^2 \frac{\kappa_2}{\Lambda_{\chi}^2} tr
({\cal M}^{\dagger} {\cal M} )\  \} \nonumber
\end{eqnarray}
with $D_{\mu} U=\partial_{\mu} U-ir_{\mu} U + iU l_{\mu}$, and
${\cal M}=\mbox{diag}( m_u,m_d,m_s)$. $l_{\mu}$ and $r_{\mu}$ are
left- and right-handed gauge fields, respectively. The unitary
matrix $U$ is a non-linear representation of the pseudoscalar
meson nonet given as $U= e^{i\Pi /f}$ with $\Pi=\pi^a\lambda_a$
and $tr(\lambda_a\lambda_b)=2\delta_{ab}$. Here we keep the
leading terms at large $N_c$ limit except the anomaly term which
arises from the order of $1/N_c$. Note that in order to make clear
 for two independent expansions, namely $1/N_c$ expansion
characterized by $p^2/\Lambda_f^2$ in the large $N_c$ limit, and
the momentum expansion described by $p^2/\Lambda_{\chi}^2$, it is useful
to introduce the scaling factor $\Lambda_{\chi} \simeq 1$ GeV and
to redefine the low energy coupling constants $L_i$ introduced in
ref. \cite{CHPT} via $L_i = \chi_i\ f^2/4\Lambda_{\chi}^2 $ and
$H_j = \kappa_j\ f^2/4\Lambda_{\chi}^2$, so that the coupling
constants $\chi_i$ ($i=3,5,8$) and $\Lambda_{\chi}$ are constants
in the large $N_c$ limit and the whole Lagrangian is multiplied by
$f^2$ and is of order $N_c$ except the U(1) anomalous term. Numerically,
one sees that $\chi_i = O(1)$ for $\Lambda_{\chi} = 1$ GeV.

 with the above chiral Lagrangian, the quark currents and densities can be represented
 by the chiral fields
 \begin{eqnarray}
\bar{q}_{jL}\gamma^{\mu} q_{iL} & \equiv &
\frac{\delta {\cal L} }{\delta(l_{\mu}(x))_{ji}}\,= -
i\frac{f^2}{2}\{ U^{\dagger}\partial^{\mu} U  \nonumber \\
& & -  r\frac{\chi_5}{2\Lambda_{\chi}^2} \left(\partial^{\mu} U^{\dagger}{\cal M}
-{\cal M}^{\dagger}\partial^{\mu} U
+\partial^{\mu} U^{\dagger} U {\cal M}^{\dagger} U - U^{\dagger}{\cal M} U^{\dagger}
\partial^{\mu} U \right) \}_{ij}\ , \\
\bar{q}_{jR} q_{iL}
& \equiv & -\frac{\delta {\cal L} }{\delta{\cal M}_{ji}}
= -r\frac{f^2}{4}\left( U^{\dagger} + \frac{\chi_5}{\Lambda_{\chi}^2}
 \partial_{\mu} U^{\dagger} \partial^{\mu} U  U^{\dagger} +
 2r \frac{\chi_8}{\Lambda_{\chi}^2} U^{\dagger}{\cal M} U^{\dagger} +
 r \frac{\kappa_2}{\Lambda_{\chi}^2} {\cal M}^{\dagger} \right)_{ij}
\end{eqnarray}
Similarly one can obtain the right-handed currents and densities. With these definitions,
all the current $\times$ current and density $\times $ density four quark operators
can be reexpressed in terms of the chiral fields, we may call such chiral representations of
four quark operators $Q_i$ as chiral operators denoted by $Q^{\chi}_i$
correspondingly.

  In the standard model, it has been found that the $\Delta S = 1$ low energy ($\mu < m_c$)
  effective Hamiltonian for calculating $K\rightarrow \pi \pi$ decay amplitudes
  can be written as

\begin{equation}
{\cal H}_{eff}^{ \Delta S=1}= \frac{G_F}{\sqrt{2}}
\;\lambda_u\sum_{i=1}^8 c_i(\mu)\,Q_i(\mu)\  ,
 \quad ( \mu < m_c)
\end{equation}
with $Q_i$ the four quark operators. Note that only seven operators are independent as the
linear relation $Q_4 = Q_2 - Q_1 + Q_3$.  $Q_7$ and $Q_8$
originate from electroweak penguin diagrams. $c_i(\mu)$ are Wilson
coefficient functions $c_i(\mu)=z_i(\mu)+\tau y_i(\mu)$.
Where $\tau=-\lambda_t/\lambda_u $ with $\lambda_q=V_{qs}^*\,V_{qd}$.
The hard task is for calculating the hadronic matrix elements $
\langle Q_i(\mu) \rangle_I $ for $\mu < \Lambda_{\chi}= 1$ GeV
which is at the order of chiral symmetry breaking scale. This is
because perturbative QCD becomes unreliable in such a low energy
scale.  It has been shown\cite{g1} that the ChPT with functional cut-off
momentum could provide a powerful way to evaluate $ \langle Q_i(\mu) \rangle_I $
when $\mu < \Lambda_{\chi}$. The procedure is as follows:
Firstly one represents the current $\times$ current or density $\times$
density four quark operators $Q_i$ by bosonized chiral fields from
the chiral Lagarangian, i.e., $Q_i^{\chi}$, then calculate loop contributions by using
 the functional cut-off momentum scheme. Finally, one matches the
two results obtained from QCD and ChPT with functional cut-off
momentum by requiring scale independence of the physical results.
$Q^{\chi}_i$ can be written as the following form\cite{g1}
\begin{eqnarray}
Q_1^{\chi} + H.c. & = & - f^4\ tr\left(\lambda_6
U^{\dagger}\partial_{\mu} U \right) tr \left(\lambda^{(1)}
U^{\dagger}\partial^{\mu} U \right) + O(1/\Lambda_{\chi}^2 ) \ ,
\nonumber \\ Q_2^{\chi} + H.c. & = & - f^4\ tr\left(\lambda_6
U^{\dagger}\partial_{\mu} U \lambda^{(1)} U^{\dagger}
\partial^{\mu} U\right) + O(1/\Lambda_{\chi}^2 )  \ ,  \nonumber
\\ Q_3^{\chi} + H.c. & = & - f^4\ tr\left(\lambda_6
U^{\dagger}\partial_{\mu} U \right) tr
\left(U^{\dagger}\partial^{\mu} U \right) + O(1/\Lambda_{\chi}^2 )
\ , \nonumber \\ Q_4^{\chi} + H.c. & = & - f^4\ tr\left(\lambda_6
\partial_{\mu} U^{\dagger} \partial^{\mu} U\right) +
O(1/\Lambda_{\chi}^2 ) \ ,  \nonumber \\ Q_5^{\chi} + H.c. & = & -
f^4\ tr\left(\lambda_6 U^{\dagger}\partial_{\mu} U \right) tr
\left(U\partial^{\mu} U^{\dagger} \right) + O(1/\Lambda_{\chi}^2 )
\ , \\ Q_6^{\chi} + H.c. & = & + f^4\
\left(\frac{r^2\chi_5}{\Lambda_{\chi}^{2}}\right)
tr\left(\lambda_6 \partial_{\mu} U^{\dagger} \partial^{\mu}
U\right) + O(1/\Lambda_{\chi}^4 )  \ ,  \nonumber \\ Q_7^{\chi} +
H.c. & = & -\frac{1}{2}Q_5^{\chi}- \frac{3}{2}f^4\
tr\left(\lambda_6 U^{\dagger}\partial_{\mu} U \right) tr
\left(\lambda^{(1)} U\partial^{\mu} U^{\dagger} \right) +
O(1/\Lambda_{\chi}^2 ) \ , \nonumber \\ Q_8^{\chi} + H.c. & = &
-\frac{1}{2} Q_6^{\chi} + f^4 r^{2}\frac{3}{4}\, tr\left(\lambda_6
U^{\dagger} \lambda^{(1)} U \right) \nonumber \\
 &+&f^4 r^{2}\frac{3}{4}\frac{\chi_5}{\Lambda_{\chi}^{2}} tr\lambda_6
 \left(U^{\dagger}\lambda^{(1)}U \partial_{\mu}U^{\dagger}\partial^{\mu} U
 +\partial_{\mu} U^{\dagger}\partial^{\mu}U
 U^{\dagger}\lambda^{(1)}U\right) \nonumber \\
 &+& f^4 r^{2}\frac{3}{4}\frac{\chi_8 }{\Lambda_{\chi}^{2}}\, 2r\,
 tr\lambda_6 \left( U^{\dagger}\lambda^{(1)} U {\cal M}^{\dagger} U
 +U^{\dagger}{\cal M}
U^{\dagger}\lambda^{(1)}U \right)
+O(1/\Lambda_{\chi}^4).\nonumber
 \end{eqnarray}
with the matrix $\lambda^{(1)}$=diag.(1,0,0). Thus loop
contributions of the chiral operators $Q^{\chi}_i$ can be
systematically calculated by using ChPT with functional cut-off
momentum.

 For $K\rightarrow \pi \pi$ decay amplitudes and direct CP-violating parameter
$\varepsilon'/\varepsilon$, the most important chiral operators
are $Q^{\chi}_1$, $Q^{\chi}_2$, $Q^{\chi}_6$ and $Q^{\chi}_8$. In
fact, the chiral operators $Q^{\chi}_3$ and $Q^{\chi}_5$ decouples
from the loop evaluations at the $p^2$ order, i.e.
\begin{equation}
Q_5^{\chi} =  Q_3^{\chi}  = 0
\end{equation}
which can explicitly be seen from the above chiral representations
due to the traceless factor $tr \left(U\partial^{\mu} U^{\dagger}
\right) = 0$ when ignoring the singlet U(1) nonet term which is
irrelevant to the Kaon decays. Here $U\partial^{\mu} U^{\dagger} =
A_{\mu}^a \lambda^a$ may be regarded as a pure gauge. As a consequence,
it implies that at the lowest
order of $p^2$, we arrive at two additional algebraic chiral
relations
 \begin{equation}
Q_4^{\chi} = Q_2^{\chi} - Q_1^{\chi}  = -f^4\,
tr\left(\lambda_6 \partial_{\mu} U^{\dagger} \partial^{\mu} U\right)
+ O(1/\Lambda_{\chi}^2) \,.
\end{equation}
and
\begin{equation}
Q_6^{\chi}  =  - \left(\frac{r^2\chi_5}{\Lambda_{\chi}^{2}}\right)
\left( Q_2^{\chi} - Q_1^{\chi} \right) = \left(\frac{r^2\chi_5}{\Lambda_{\chi}^{2}}\right)
f^4\,
tr\left(\lambda_6 \partial_{\mu} U^{\dagger} \partial^{\mu} U\right)
\end{equation}
Notice that the mass parameter $r$ is at the same order of the
energy scale $\Lambda_{\chi}$, and $\chi_5$ is at order of unit,
thus the leading non-zero contribution of $Q_6^{\chi}$ is at the
same order of  $Q_2^{\chi}$ and $Q_1^{\chi}$.

The above algebraic chiral relations mentioned also as Wu-relations
 in the literature \cite{BJL} were first derived in
ref.\cite{YLW1}, they have been checked from an explicit
calculation up to the chiral one-loop level by using the usual
cut-off regularization\cite{BBG}. It was based on this observation that
the ratio $\varepsilon'/\varepsilon$ was predicted\cite{YLW1,PW} to be large enough
 ( $\varepsilon'/\varepsilon = (1-3)\times 10^{-3}$ ) for observation.
 In ref. \cite{g1}, it has further been shown that the algebraic Wu-relations
 of the chiral operators survive from loop corrections and
 can provide a consistent prediction for both the direct CP-violating parameter
 $\varepsilon'/\varepsilon$ and the $\Delta I = 1/2$ rule when applying for an
 appropriate matching procedure between QCD and ChPT.

\section{Matching Approach}

 The short-distance evolution of quark operators is performed from perturbative QCD.
 When the energy scale $\mu$ is high, $m_W > \mu > m_b$, there are eleven
 independent operators $Q_i$ ($i=1,\cdots, 11$). When the energy scale $\mu$
 runs down to below the bottom quark mass $m_b$ and above the charm
 quark mass $m_c$, i.e., $m_b > \mu > m_c$, the operator $Q_{11}$ decouples and
 operator $Q_{10}$ is given by the linear combination $Q_{10} = -2Q_1 + 2Q_2
 + Q_3 - Q_4$. Once the energy scale $\mu$
 goes down to below $m_c$ but above the confining scale or the energy
 scale $\Lambda_{\chi}$, i.e., $m_c > \mu > \Lambda_{\chi}$, two
 operators $Q_9$ and $Q_4$ become no longer independent and are given by the
 linear combination $Q_9 = Q_2 + Q_1$ and $Q_4 = Q_3 + Q_2 - Q_1$. Thus there
 are only seven independent operators below $m_c$ and above $\Lambda_{\chi}$.
 The one-loop QCD corrections of the quark operators at
 the energy scale just above the energy scale $\Lambda_{\chi}$ is given by
\begin{eqnarray}
Q_1(\mu_Q) & = & Q_1(\mu) - 3 \frac{\alpha_s}{4\pi}\ln (\frac{\mu_Q^2}{\mu^2})\,
Q_2(\mu) + O(1/N_c)\,   , \\
Q_2(\mu_Q) & = & Q_2(\mu) - 3 \frac{\alpha_s}{4\pi}\ln (\frac{\mu_Q^2}{\mu^2})\,
Q_1(\mu) \nonumber \\
 & - & \frac{1}{3}\, \frac{\alpha_s}{4\pi}\ln (\frac{\mu_Q^2}{\mu^2})\, Q_4(\mu)
- \frac{1}{3}\, \frac{\alpha_s}{4\pi}\ln (\frac{\mu_Q^2}{\mu^2})\, Q_6(\mu) + O(1/N_c)\,  , \\
Q_4(\mu_Q) & = & Q_4(\mu) - 3 \frac{\alpha_s}{4\pi}\ln (\frac{\mu_Q^2}{\mu^2})\,
Q_3(\mu) \nonumber \\
& - & \frac{\alpha_s}{4\pi}\ln (\frac{\mu_Q^2}{\mu^2})\, Q_4(\mu)
-  \frac{\alpha_s}{4\pi}\ln (\frac{\mu_Q^2}{\mu^2})\, Q_6(\mu) + O(1/N_c)\, , \\
Q_6(\mu_Q) & = & Q_6(\mu)
- \frac{\alpha_s}{4\pi}\ln (\frac{\mu_Q^2}{\mu^2})\, Q_4(\mu)
\nonumber \\
 & + & [3(N_c -1/N_c)- 1] \frac{\alpha_s}{4\pi}\ln (\frac{\mu_Q^2}{\mu^2})\,  \, Q_6(\mu)
+ O(1/N_c)\, ,\\
Q_8(\mu_Q) & = & Q_8(\mu)  +
[3(N_c -1/N_c)- 1] \frac{\alpha_s}{4\pi}\ln (\frac{\mu_Q^2}{\mu^2})\,  \, Q_8(\mu) \  ,
\end{eqnarray}
and
\begin{eqnarray}
Q_3(\mu_Q) & = & Q_3(\mu)  - \frac{11}{3}\, \frac{\alpha_s}{4\pi}
\ln (\frac{\mu_Q^2}{\mu^2})\, Q_4(\mu) - \frac{2}{3}\, \frac{\alpha_s}{4\pi}
\ln (\frac{\mu_Q^2}{\mu^2})\, Q_6(\mu) + O(1/N_c)\, , \\
Q_5(\mu_Q) & = & Q_5(\mu)  + 3\frac{\alpha_s}{4\pi}
\ln (\frac{\mu_Q^2}{\mu^2})\, Q_6(\mu) + O(1/N_c)\ ,  \\
Q_7(\mu_Q) & = & Q_7(\mu)  + 3\frac{\alpha_s}{4\pi}
\ln (\frac{\mu_Q^2}{\mu^2})\, Q_8(\mu) + O(1/N_c)\ .
\end{eqnarray}
It is then clear that

i) In the large $N_c$ limit, $Q_1$, $Q_2$, $Q_4$, and $Q_6$ form
a complete set of operators under QCD corrections.

ii) The evolution of $Q_8$ is independent of other operators and only caused
by loop corrections of the density.

iii) The operator $Q_3$ is given by the linear combination $Q_3 = Q_4 -(Q_2 -Q_1)$.
The operator $Q_5$ is driven by the operator $Q_6$, and the operator $Q_7$
is driven by the operator $Q_8$.

When the energy scale $\mu$ approaches to the confining scale,
or $\mu < \Lambda_{\chi}\sim \Lambda_F \sim 1$GeV,
as we have discussed in the above sections,
long-distance effects have to be considered. The evolution of the
operators $Q_i(\mu)$ when $\mu < \Lambda_{\chi}$ is supposed
to be carried out by the one of the chiral operators
$Q_i^{\chi}(M(\mu))$ in the framework of the functional cut-off ChPT truncated to the
pseudoscalars. To be treated at the same approximations made in the short-distance operator
evolution of QCD, we should only keep the leading terms (i.e., quadratic terms of
functional cut-off momentum) and take the chiral limit, i.e., $m_K^2, m_{\pi}^2 << \Lambda_F^2$.
The evolution of the operators $Q_1^{\chi}$ and $Q_2^{\chi}$
is simply given by
\begin{eqnarray}
Q_1(\mu)& \rightarrow & Q_1^{\chi}(M(\mu))= Q_1^{\chi}(M(\mu'))
-\frac{2(M^2(\mu)-M^2(\mu'))}{\Lambda_F^2}
\, Q_2^{\chi}(M(\mu'))\, ,\\
Q_2(\mu)& \rightarrow & Q_2^{\chi}(M(\mu))= Q_2^{\chi}(M(\mu'))
-\frac{2(M^2(\mu)-M^2(\mu'))}{\Lambda_F^2}
\, Q_1^{\chi}(M(\mu')) \nonumber \\
& + & \frac{M^2(\mu)-M^2(\mu')}{\Lambda_F^2}\, ( Q_2^{\chi} - Q_1^{\chi})(M(\mu'))\, ,
\end{eqnarray}
where $\Lambda_F = 4\pi F =1.16$ GeV with $F$ the renormalized one of $f$.
The operators $Q_i^{\chi}$ ($i=4,6,8$) can
be written as follows
\begin{eqnarray}
Q_4(\mu)& \rightarrow & Q_4^{\chi}(M(\mu)) = (Q_2^{\chi} - Q_1^{\chi})(M(\mu))  \,  ,\\
Q_6(\mu)& \rightarrow & Q_6^{\chi}(\mu, M(\mu))= \left(1 +
 3(N_c -1/N_c)\frac{\alpha_s}{4\pi}\ln (\frac{\mu^{2}}{\mu_{\chi}^{2}}) \right)\,
Q_6^{\chi}(\mu_{\chi}, M(\mu))\, , \\
Q_8(\mu)& \rightarrow & Q_8^{\chi}(\mu, M(\mu))= \left(1 +
 3(N_c -1/N_c)\frac{\alpha_s}{4\pi}\ln (\frac{\mu^{2}}{\mu_{\chi}^{2}}) \right)\,
Q_8^{\chi}(\mu_{\chi}, M(\mu))\, .
\end{eqnarray}
where the explicit $\mu$-dependence of the operators
$Q_6^{\chi}(\mu, M(\mu))$ and $Q_8^{\chi}(\mu, M(\mu))$ arise from
the running quark mass and behavors like $1/(m_s(\mu) +
\hat{m}(\mu))^{2}$. Notice that the operators $Q_3^{\chi}$ and
 $Q_5^{\chi}$ decouple from the evolution, namely
$Q_3^{\chi}=0$ and $Q_5^{\chi}=0$. Thus the independent operators are
reduced once more in the long-distance operator evolution when
$\mu < \Lambda_{\chi}$ due to the algebraic chiral operator
relations.

 Matching the loop results evaluated
from QCD with the ones from the ChPT with functional cut-off
momentum at the energy scale $\Lambda_{\chi}$ with large $N_c$ expansion
and requiring $\mu$-independence in the large $N_c$ limit, i.e.,
$\frac{\partial}{\partial \mu} Q_1(\mu_Q)=0$,
we arrive at the two matching conditions\cite{g1},
\begin{equation}
\mu \frac{\partial}{\partial \mu} \left( \frac{2M^2(\mu)}{\Lambda_F^2}\right)
= \frac{3\alpha_s}{2\pi}\, ,
\end{equation}
\begin{equation}
Q_6^{\chi}(\mu_{\chi}, M(\mu)) =  -\frac{11}{2}(Q_2^{\chi} - Q_1^{\chi} ) (M(\mu)) \, ,
\quad \mu < \Lambda_{\chi}
\end{equation}

When combining the second matching condition with the chiral Wu-relation
\begin{eqnarray}
& & Q_6^{\chi}(\mu_{\chi}, M(\mu)) \simeq
\left(- \frac{R^2_{\chi} \chi_5 ^r}{\Lambda_{\chi}^2}\right)
( Q_2^{\chi} - Q_1^{\chi})(M(\mu))\, , \quad \mu < \Lambda_{\chi} \\
& & R_{\chi} \equiv R(\mu\simeq \mu_{\chi} )\simeq m_{\pi}^{2}/\hat{m}(\mu_{\chi})
\simeq 2m_{K}^{2}/(m_s + \hat{m})(\mu_{\chi})\,  ,
\end{eqnarray}
 we are able to fix the strange quark mass\cite{g1}
\begin{equation}
\frac{R^2_{\chi} \chi_5 ^r}{\Lambda_{\chi}^2}  = \frac{11}{2}\,  \quad \rightarrow \quad
m_s(\mu_{\chi}) \simeq 196 MeV\, ,
\end{equation}
Note that the coupling constants $\chi_5$ and $r$ must be replaced by the corresponding
renormalized ones $\chi_5 ^r$ and $R(\mu)$ as their loop corrections are
at the subleading order. Here we have also used the result $\Lambda_{\chi} =
1.03 \sqrt{\chi_5^r}$ GeV which is fixed from the ratio of the kaon and
pion decay constants.

 After integrating the first matching condition, we find that
 the $\mu$-dependence of the functional cut-off momentum $M(\mu)$ can
be written as\cite{g1}
\begin{equation}
\frac{2M^2(\mu)}{\Lambda_F^2} \simeq \frac{3\alpha_s}{4\pi}
 + \frac{3\alpha_s}{4\pi}\, \ln(\frac{\mu^2}{\mu_0^2} )\, ,
 \end{equation}
which shows that after imposing the matching condition for the
anomalous dimensions between quark operators $Q_i(\mu)$ in QCD and
the corresponding chiral operators $ Q_i^{\chi}(M(\mu))$ in the
ChPT with functional cut-off momentum, the dimensionless ratio
$M^2/\Lambda_F^2$ is only related to the strong coupling constant
$\alpha_s$ and becomes scheme-independent, which implies that the
long-distance operator evolution in  ChPT with functional cut-off
momentum can be carried out by using any approach. Where $\mu_0 $
is the low energy scale arised as the integrating constant.
In general, we have $\mu_0 > \Lambda_{QCD}$. To fix the value of
$\mu_0$, we use $M_0 \simeq \mu_0$. Thus $\mu_0$ (or
$\alpha_s(\mu_0)$ ) is determined via
$\mu_0 \simeq \Lambda_F \sqrt{3\alpha_s(\mu_0)/8\pi}$.
 Using the definition $\alpha_s(\mu) = 6\pi /[(33-2n_f)
 \ln (\mu^2/\Lambda_{QCD}^2)]$ with $n_f =3$,
 $\mu_0$ is found to be (for $\Lambda_{QCD} = 325\pm 80$ MeV)
\begin{equation}
 \mu_0 \simeq 435\pm 70 MeV
  \quad or \quad  \alpha_s(\mu_0)/2\pi \simeq 0.19^{+0.06}_{-0.05}     \  .
\end{equation}
Thus the functional cut-off
momentum $M(\mu)$ at $\mu = \Lambda_{\chi}$ yields the following
corresponding value
\begin{equation}
M_{\chi} \equiv M(\mu = \Lambda_{\chi} \simeq 1 GeV) \simeq
0.71^{+0.11}_{-0.12} GeV \ .
\end{equation}
which provides the possible allowed range of the energy scale
where the ChPT with functional cut-off momentum can be used to
describe the low energy behavior of QCD at large $N_c$ limit.

\section{Long-distance Evolution of Chiral Operators}

 We may now adopt the matching conditions and algebraic Wu-relations of
 the chiral operators to investigate the long-distance operator evolution.
 The $\Delta S = 1$ low energy ($\mu < \Lambda_{\chi}$)
 effective Hamiltonian for calculating $K\rightarrow \pi \pi$ decay amplitudes may be
 expressed as follows
\begin{equation}
{\cal H}_{eff}^{ \Delta S=1}= \frac{G_F}{\sqrt{2}}
\;\lambda_u\sum_{i=1,2,4,6,8} c_i(\Lambda_{\chi})\,Q_i^{\chi}(M(\Lambda_{\chi}))\
\end{equation}
 It is seen that the first matching condition enables us to sum over all the leading terms via
 renormalization group equation down to the energy scale $\mu_0$, and the second
  matching condition together with
 the algebraic chiral operator relations allows us to evaluate the penguin operators
$Q_4^{\chi}(M)$ and $Q_6^{\chi}(M)$ from the operators $Q_1^{\chi}(M)$ and $Q_2^{\chi}(M)$.
So that the operators $Q_1^{\chi}(M)$ and $Q_2^{\chi}(M)$ form a complete set for the operator
evolution below the energy scale $\mu \simeq \Lambda_{\chi} \simeq 1 $ GeV, or correspondingly,
 below the functional cut-off momentum $M(\mu\simeq \Lambda_{\chi})
  \simeq 0.71^{+0.11}_{-0.12} $ GeV for $\Lambda_{QCD} = 325 \pm 80$ MeV.
 By choosing a new operator basis
 $Q_{\pm}^{\chi}(M(\mu)) = Q_2^{\chi}(M(\mu))\pm Q_2^{\chi}(M(\mu))$ with
 the anomalous dimension matrix for the
 basis $(Q_{-}^{\chi}, Q_{+}^{\chi})$
 \begin{equation}
 \gamma = \frac{\alpha_{s}}{2\pi} \left( \begin{array}{cc}
 -9/2 & 0 \\
 -3/2 & 3 \\
 \end{array} \right)
 \end{equation}
  and following the standard procedure of the renormalization group evolution
   with the initial conditions for the Wilson coefficient functions:
  $c_{-}(\Lambda_{\chi}) = c_2 (\Lambda_{\chi}) - c_1 (\Lambda_{\chi})$ and
  $c_{+}(\Lambda_{\chi}) = c_2 (\Lambda_{\chi}) + c_1(\Lambda_{\chi}) $ ,
  we find in the leading logarithmic approximation that
  \begin{eqnarray}
  Q_{-}^{\chi}(M(\Lambda_{\chi})) & = & \eta_{\chi} ^{-1/2}\,
  \eta_{-}(M_{\chi})\, Q_{-}^{\chi}(\mu_0)\ , \\
  Q_{+}^{\chi}(M(\Lambda_{\chi})) & = &
  \eta_{\chi}^{1/3}\, \eta_1(M_{\chi})\, Q_{+}^{\chi}(\mu_0) + \frac{1}{5}\left(
  \eta_{\chi}^{-1/2} - \eta_{\chi}^{1/3} \right)\, \eta_2(M_{\chi})\, Q_{-}^{\chi}(\mu_0)\  ,
  \end{eqnarray}
   with $\eta_{\chi} = \alpha_{s}(\Lambda_{\chi})/\alpha_{s}(\mu_0)$, and
  \begin{eqnarray}
  Q_{-}^{\chi}(\mu_0) & = & Q_{-}^{\chi}(0) + \frac{9\alpha_s(\mu_0)}{8\pi}\
  Q_{-}^{\chi}(0)\  , \\
  Q_{+}^{\chi}(\mu_0) & = & Q_{+}^{\chi}(0) -  \frac{3\alpha_s(\mu_0)}{4\pi}\
  Q_{+}^{\chi}(0) +  \frac{3\alpha_s(\mu_0)}{8\pi}\  Q_{-}^{\chi}(0)\  ,
  \end{eqnarray}
 where $\eta_{-}(M_{\chi})$, $\eta_{1}(M_{\chi})$ and $\eta_{2}(M_{\chi})$
  represent the finite meson mass contributions
\begin{eqnarray}
\eta_{-}(M_{\chi})& \simeq & 1 + \frac{\frac{3}{4}m_K^2 -
\frac{9}{2}m_{\pi}^2}{\Lambda_{F}^2}
 \ln \left(1 + \frac{M^2(\mu)}{\tilde{m}^2}\right) \, , \nonumber \\
\eta_{1}(M_{\chi})& \simeq & 1 + \frac{\frac{1}{4}m_K^2 +
3m_{\pi}^2}{\Lambda_{F}^2}
 \ln \left(1 + \frac{M^2(\mu)}{\tilde{m}^2}\right) \, , \nonumber \\
 \eta_{2}(M_{\chi})& \simeq & 1 + \frac{m_K^2 -
\frac{3}{2}m_{\pi}^2}{M_{\chi}^2}
 \ln \left(1 + \frac{M^2(\mu)}{\tilde{m}^2}\right) \, .
\end{eqnarray}
Numerically, we use $\tilde{m}\simeq 300$MeV, $m_{K} = 0.495$ GeV
and $m_{\pi} = 0.137$GeV. When the QCD scale takes the value
$\Lambda_{QCD} = 325\pm 80$ MeV with the corresponding
  low energy cut-off momentum $\mu_0 \simeq 435\pm 70$ MeV, we have\cite{g1}
  \begin{eqnarray}
  Q_{-}^{\chi}(M(\Lambda_{\chi})) & = & (3.17^{+0.66}_{-0.43})\  Q_{-}^{\chi}(0)
  =  Q_{4}^{\chi}(M(\Lambda_{\chi}))\  ,  \\
  Q_{+}^{\chi}(M(\Lambda_{\chi})) & = & (0.55^{-0.09}_{+0.06})\ Q_{+}^{\chi}(0)
  + (0.8^{+0.11}_{-0.05})\ Q_{-}^{\chi}(0)\  , \\
  Q_{6}^{\chi}(\mu_{\chi}, M(\Lambda_{\chi})) & = & -\frac{11}{2}
   Q_{-}^{\chi}(M(\Lambda_{\chi}))
   = - (17.44^{+3.62}_{-2.37})\  Q_{-}^{\chi}(0)\  , \\
  Q_{8}^{\chi}(\mu_{\chi}, M(\Lambda_{\chi})) & = & \frac{33}{8}\
  \frac{\Lambda_{\chi}^{2}}{\chi_5^{r}(m_K^2 - m_{\pi}^2 )}\
  ( Q_{+}^{\chi} + Q_{-}^{\chi} )(0) =
  19.18\ ( Q_{+}^{\chi} + Q_{-}^{\chi} )(0)\ .
  \end{eqnarray}
   with these analyzes, we are able to present our numerical predictions
   for the isospin amplitudes and the direct CP-violating parameter
    $\varepsilon'/\varepsilon$.

    \section{ New Predictions for $\varepsilon'/\varepsilon$  and $\Delta I = 1/2$ Rule}

 The direct CP violation $\varepsilon'/\varepsilon$ in kaon decays arises from the
 nonzero relative phase of isospin amplitudes $A_0$ and $A_2$
  \begin{equation}
  \frac{\varepsilon'}{\varepsilon} = \frac{\omega}{\sqrt{2}|\varepsilon|}
  \left( \frac{Im A_2}{Re A_2} -
  \frac{Im A_0}{Re A_0} \right)
  \end{equation}
 with $\omega = Re A_2/ Re A_0 = 1/22.2$, we arrive at the following general expression
 \begin{equation}
  \frac{\varepsilon'}{\varepsilon} = \frac{G_F}{2}\frac{\omega}{|\varepsilon|Re A_0 }
  Im \lambda_t \left( h_0 - h_2/\omega \right)
  \end{equation}
 where $A_I$ are the $K\rightarrow \pi \pi$ decay amplitudes  with isospin $I$ with
 \begin{equation}
 A_I \cos\delta_{I} = \langle \pi\pi | {\cal H}_{ef\ f}^{ \Delta S=1} | K \rangle \equiv
 \frac{G_F}{\sqrt{2}} \;\lambda_u\sum_{i=1,2,4,6,8} c_i(\Lambda_{\chi})
 Re\langle Q_i^{\chi}(M(\Lambda_{\chi}))\rangle_I
 \end{equation}
 The CP-conserving amplitudes are given by
 \begin{eqnarray}
 Re A_0 \cos\delta_{0}& = & \frac{G_F}{\sqrt{2}} \; Re\lambda_u\sum_{i=1,2,4,6,8}
 z_i(\Lambda_{\chi})Re\langle Q_i^{\chi}(M(\Lambda_{\chi}))\rangle_0 \nonumber \\
 & \simeq & \frac{G_F}{\sqrt{2}} \; Re\lambda_u [
 \frac{1}{2}z_{-}(\Lambda_{\chi})Re\langle Q_{-}^{\chi}(M(\Lambda_{\chi}))\rangle_0 +
 \frac{1}{2}z_{+}(\Lambda_{\chi})Re\langle Q_{+}^{\chi}(M(\Lambda_{\chi}))\rangle_0 \nonumber \\
 & + & z_{4}(\Lambda_{\chi})Re\langle Q_{4}^{\chi}(M(\Lambda_{\chi}))\rangle_0 +
 z_{6}(\Lambda_{\chi})Re\langle Q_{6}^{\chi}(M(\Lambda_{\chi}))\rangle_0 ] \  , \\
 Re A_2 \cos\delta_{2} & = & \frac{G_F}{\sqrt{2}} \; Re\lambda_u\sum_{i=1,2,8}
 z_i(\Lambda_{\chi})Re\langle Q_i^{\chi}(M(\Lambda_{\chi}))\rangle_2 \\
 & \simeq & \frac{G_F}{\sqrt{2}} \; Re\lambda_u [
 \frac{1}{2}z_{-}(\Lambda_{\chi})Re\langle Q_{-}^{\chi}(M(\Lambda_{\chi}))\rangle_2 +
 \frac{1}{2}z_{+}(\Lambda_{\chi})Re\langle Q_{+}^{\chi}(M(\Lambda_{\chi}))\rangle_2 ]
  \  , \nonumber
 \end{eqnarray}
 and the hadronic matrix elements concerned for calculating the ratio
 $\varepsilon'/\varepsilon$ are
  \begin{eqnarray}
  h_0 & = & (\cos\delta_0)^{-1}\sum_{i=1,2,4,6,8} y_i(\Lambda_{\chi})
  Re\langle Q_i^{\chi}\left(M(\Lambda_{\chi})\right)\rangle_0
  \left(1 - \Omega_{IB} \right) \nonumber \\
  & \simeq & (\cos\delta_0)^{-1} y_6(\Lambda_{\chi})
  Re\langle Q_6^{\chi}\left(M(\Lambda_{\chi})\right)\rangle_0
   \left(1 - \Omega_{IB} \right)\  , \\
  h_2 & = & (\cos\delta_2)^{-1}\sum_{i=1,2,8} y_i(\Lambda_{\chi})
  Re\langle Q_i^{\chi}\left(M(\Lambda_{\chi})\right)\rangle_2  \nonumber \\
  &\simeq & (\cos\delta_2)^{-1} y_8(\Lambda_{\chi})
  Re\langle Q_8^{\chi}\left(M(\Lambda_{\chi})\right)\rangle_2 \  ,
  \end{eqnarray}
 where we have taken into account the possible isospin breaking effect
$\Omega_{IB}$ with the most recent refined
calculation in\cite{ISB1}: $\Omega_{IB}
\simeq 0.16 \pm 0.03$, which is smaller than the previously estimated value $\Omega_{IB}
\simeq 0.25 \pm 0.1$\cite{ISB}, but with a large error\cite{ISB2}.
The CKM factors $Re\lambda_u$ and $Im\lambda_t$ are given in the
Wolfenstein parameterization\cite{LW} as follows
 \begin{equation}
 Re\lambda_u = Re(V_{us}^{\ast}V_{ud})= \lambda\ , \quad
Im\lambda_t = Im(V_{ts}^{\ast}V_{ta}) = A^2\lambda^5 \eta
 \end{equation}

  To obtain the numerical results, we use the following reliable values for all
  relevant parameters: $\Lambda_{QCD} = 325\pm 80$ MeV,
  $\mu_{0} = 435\pm 70 $MeV, $\Lambda_{\chi}=1.0$GeV and $\Lambda_F = 1.16$ GeV;
  $z_{-}(\Lambda_{\chi}) =(z_2 - z_1)(\Lambda_{\chi}) = 2.181^{+0.197}_{-0.177}$,
  $z_{+}(\Lambda_{\chi}) = (z_2 + z_1)(\Lambda_{\chi}) = 0.685\mp 0.029$,
  $z_{4}(\Lambda_{\chi}) = -(0.012\pm 0.003)$ and $z_{6}(\Lambda_{\chi}) = -(0.013\pm 0.003)$,
  as well as $y_{6}(\Lambda_{\chi}) = -\left(0.113^{+0.024}_{-0.021}\right)$
  and $y_{8}(\Lambda_{\chi})/\alpha = 0.158^{+0.040}_{-0.033}$;
  $\langle Q_{-}^{\chi}(0)\rangle_0 = 36.9\times 10^6 $ MeV$^{3}$,
  $\langle Q_{+}^{\chi}(0)\rangle_0 = 12.3\times 10^6 $ MeV$^{3}$,
  $\langle Q_{+}^{\chi}(0)\rangle_2 = 34.8\times 10^6 $ MeV$^{3}$ and
  $\langle Q_{8}^{\chi}(\Lambda_{\chi}, 0)\rangle_2= 328.8\times 10^6 $ MeV$^{3}$.
  Note that for the
  Wilson coefficient functions, we only use the leading order results at one-loop level
  for a consistent analysis since the chiral operators have only been evaluated up to
  the leading order at the chiral one-loop level, namely at the order of $1/N_c\sim
  M^2/\Lambda_{F}^2\sim \alpha_s$ in the large $N_c$ approach.
  Their values can be read following the calculations in refs. \cite{LO}.
  The numerical values at $\mu = \Lambda_{\chi}$ are regarded as the `initial conditions'
  for the chiral operator evolution and read for $\Lambda_{QCD} = 325\pm 80$ MeV.
  For the hadronic matrix
  elements of chiral operators at cut-off momentum $M=0$ take their values at the tree-level.
  For the CKM matrix elements, there remain big uncertainties arising from the
  single CP-violating phase, two matrix elements $V_{ub}$ and $V_{cb}$,
  or the corresponding Wolfenstein parameters $\eta$, $\rho$ and $A$. For a numerical estimate,
  we take $Re\lambda_u = 0.22$ and $Im\lambda_t =1.2 \times 10^{-4}$ as the central
  values\cite{LT}.
  With these input values, we obtain the isospin amplitudes
  \begin{eqnarray}
  Re A_0 & = &  (2.56^{+0.78}_{-0.37})\, \times 10^{-4}\ (\cos\delta_0)^{-1}\ MeV
  = (3.10^{+0.94}_{-0.61})\, \times 10^{-4}\ MeV \  , \\
  Re A_2 & = &  (0.12\mp 0.02) \times 10^{-4}\ (\cos\delta_2)^{-1}\ MeV
   = (0.12\mp 0.02)\times 10^{-4}\ MeV \  ,
  \end{eqnarray}
 which is consistent with the experimental data: $Re A_0 = 3.33 \times 10^{-4}$ MeV and
 $Re A_2 = 0.15 \times 10^{-4}$ MeV . Here the final state interaction phases,
 $\delta_0 = (34.2\pm 2.2)^o$ and $\delta_2 = (-6.9 \pm 0.2)^o$ \cite{CO} have been used.
 Simultaneously, it leads to a consistent prediction for
  the direct CP-violating parameter $\varepsilon'/\varepsilon$
  \begin{equation}
  \frac{\varepsilon'}{\varepsilon} = (23.6^{+12.4}_{-7.8})\, \times 10^{-4}
   \left(\frac{Im \lambda_t}{ 1.2 \times 10^{-4}} \right)
  \end{equation}

 From the above analyzes, it is noticed that

 1. The main uncertainties for the predictions arise from the QCD scale $\Lambda_{QCD}$
 (or the low energy scale $\mu_0$) and the combined CKM factor
 $Im\lambda_t$. Nevertheless, the uncertainties from the energy scale may be reduced from
 comparing the predicted isospin amplitudes $A_0$ and $A_2$ with the well measured
 ones. It is seen that the results corresponding to the large values
  of $\Lambda_{QCD} > 325$ MeV appear not favorable.

 2. Considering from the isospin amplitude $A_2$,
 the ratio $\varepsilon'/\varepsilon$ favors the low values
\begin{equation}
\frac{\varepsilon'}{\varepsilon} \simeq 16 \times 10^{-4}\ (Im \lambda_t /1.2 \times 10^{-4} )
 \end{equation}
 while from the isospin amplitude $A_0$, it favors the high values
\begin{equation}
\frac{\varepsilon'}{\varepsilon} \simeq 24\times 10^{-4}\ (Im \lambda_t /1.2 \times 10^{-4} )
 \end{equation}
 From the ratio of two amplitudes $ReA_0/ReA_2$ , i.e., the $\Delta I =1/2$
 rule, the ratio $\varepsilon'/\varepsilon$ favors the middle values
\begin{equation}
\frac{\varepsilon'}{\varepsilon} \simeq 20 \times 10^{-4} \ (Im \lambda_t /1.2 \times 10^{-4} )
 \end{equation}

 3. The above results are renormalization scheme independent
 as the consistent matching
 between QCD and ChPT  occurs at the leading one-loop
 order of $1/N_c\sim \alpha_s \sim 1/\Lambda_F^2$
 around the scale $\Lambda_{\chi}$. The renormalization scheme
 dependence arises from the next-to-leading order of perturbative QCD\cite{NTL}, which
 could become substantial for some of the Wilson coefficient functions when the
 renormalization scale $\mu$ runs down to around the scale $\Lambda_{\chi} =
 1$GeV. For the long-distance evolution, the scheme is fixed by the ChPT
 with functional cut-off momentum. For
 matching to this scheme, it is useful to introduce a scheme independent basis for the
 perturbative QCD calculation of short-distance physics. Then applying our above procedure to
 find out the matching conditions at the next-to-leading order
 $1/N_c^2\sim \alpha_s^2 \sim 1/\Lambda_F^4$. To work out the scheme independent basis in QCD,
 it may be helpful to adopt the method discussed in ref.\cite{WB} and use the cut-off
 momentum basis.

 In summary, our new consistent prediction for the direct CP-violating
 parameter $\varepsilon'/\varepsilon$ is
 \begin{equation}
\frac{\varepsilon'}{\varepsilon} = (20\pm 4) \times 10^{-4}
\left(\frac{Im \lambda_t}{1.2 \times 10^{-4} } \right) = (20\pm 4 \pm 5) \times 10^{-4}
 \end{equation}
   where the first error from the low energy scale and the second one from the CKM factor
   $Im \lambda_t$. Our prediction is consistent with the most recent results reported by the
 NA48 collaboration at CERN\cite{NA48} and the KTeV collaboration at
 Fermilab\cite{KTEV}
 \begin{eqnarray}
   Re(\varepsilon'/\varepsilon) & = & (20.7\pm 2.8)\times 10^{-4}
  \quad (2001\, KTeV) \cite{KTEV}\, \\
   Re(\varepsilon'/\varepsilon) & = & (15.3\pm 2.6)\times
   10^{-4}, \quad (2001\, NA48)\cite{NA48}\, ;
  \end{eqnarray}
as well as with the world average
  \begin{equation}
  Re(\varepsilon'/\varepsilon)  =  (17.2\pm 1.8)\times 10^{-4}
  \quad (World\, \ Average\, \ 2001)
  \end{equation}

  \section{Conclusions}

    Let me briefly summarize the new predictions by Beijing group:
   \begin{itemize}
   \item  We have Clearly made a bi-expansion: $1/N_c \sim \alpha_s \sim M^2/\Lambda_f^2
   $ and $p^2/\Lambda_{\chi}^2$ ( $m_q^2/\Lambda_{\chi}^2$) by introducing the $N_c$-independent
   energy scale. The leading non-zero contributions for relevant
   chiral operators are all at the same order of $\sqrt{N_c}$.
   \item Chiral representation of operators allows us to get some useful algebraic
   chiral relations which reduces the independent operators and
   relates the ratio $\varepsilon'/\varepsilon$ to the $\Delta I = 1/2$ rule.
   \item Chiral loops have been calculated by using the functional cutoff
   momentum (FCOM) ($M(\mu)$) scheme.
   The form of $M(\mu)$ is determined by matching between short- and long-distance
   contributions.
   \item The leading order of $1/N_c$ between chiral one loops and QCD one loops
   has been found to be well matched at large-$N_c$ limit. In this sense,
   ChPT does well describe the low energy dynamics of QCD.
   \item Two useful matching conditions have been resulted, which makes
   the predictions for the ratio $\varepsilon'/\varepsilon$ and the $\Delta I =1/2$ rule
   are insensitive to the strange quark mass.
   \item The results at the leading order of loops
    are both renormalization scale and  renormalization scheme independent.
   \end{itemize}

  The present applications of ChPT also indicate that the ChPT with chiral Lagrangian obtained
  based on the chiral flavor symmetry $SU(3)_L\times SU(3)_R$ and inspired from $1/N_c$
  approach is a consistent and useful effective theory. With the functional cutoff momentum
  scheme, the ChPT may well describe the low energy dynamics of QCD. We expect that the theoretical uncertainties
  can be further improved within the framework of ChPT. Nonetheless to say,
  the flavor symmetry has plaid an important role on flavor physics\cite{HF}.

\end{document}